\documentclass{Interspeech}
\usepackage{xcolor}
\usepackage{amsmath}
\usepackage{multirow}

\interspeechcameraready 

\title{SLASH: Self-Supervised Speech Pitch Estimation \\ Leveraging DSP-derived Absolute Pitch}

\author[affiliation={1}]{Ryo}{Terashima}
\author[affiliation={1}]{Yuma}{Shirahata}
\author[affiliation={1}]{Masaya}{Kawamura}

\affiliation{}{LY Corporation}{Japan}
\email{ryo.terashima@lycorp.co.jp, yuma.shirahata@lycorp.co.jp, kawamura.masaya@lycorp.co.jp}

\begin{document}

\maketitle
\setlength{\abovedisplayskip}{5pt}
\setlength{\belowdisplayskip}{5pt}
\setlength\floatsep{6pt}
\setlength\intextsep{6pt}
\setlength\textfloatsep{6pt}
\setlength{\dbltextfloatsep}{6pt}
\fontsize{9.0}{10.05}\selectfont

\begin{abstract}
We present SLASH, a pitch estimation method of speech signals based on self-supervised learning (SSL). To enhance the performance of conventional SSL-based approaches that primarily depend on the relative pitch difference derived from pitch shifting, our method incorporates absolute pitch values by 1) introducing a prior pitch distribution derived from digital signal processing (DSP), and 2) optimizing absolute pitch through gradient descent with a loss between the target and differentiable DSP-derived spectrograms. To stabilize the optimization, a novel spectrogram generation method is used that skips complicated waveform generation.
In addition, the aperiodic components in speech are accurately predicted through differentiable DSP, enhancing the method's applicability to speech signal processing.
Experimental results showed that the proposed method outperformed both baseline DSP and SSL-based pitch estimation methods, attributed to the effective integration of SSL and DSP.
\end{abstract}

\section{Introduction}
Pitch estimation is a fundamental task in the field of speech signal processing. It has been applied to many kinds of applications, such as text-to-speech and emotion recognition, among others~\cite{tan2021survey,khalil2019speech,sisman2020overview}. Historically, this task has been performed using digital signal processing (DSP)-based methods~\cite{mauch2014pyin, camacho2008sawtooth, talkin1995robust}, but with the advent of deep neural networks (DNNs), data-driven approaches have been proposed~\cite{riou2023pesto,gfeller2020spice,kim2018crepe,singh2021deepf0}. 

One of the prominent methods using DNNs is the supervised approach~\cite{gfeller2020spice,singh2021deepf0,subramani2024noise}. In this approach, a model is trained to estimate the fundamental frequency ($F_0$)\footnote{In general, the term $F_0$ represents a physical property, while pitch represents a perceptual property of sound. In this paper, for convenience, the term "pitch" is used interchangeably with $F_0$.
} from audio data using a simple framework where $F_0$ serves as the target label. Although this approach is effective, it faces challenges such as the difficulty in obtaining reliable target $F_0$ annotations and the tendency of the inference results to be overly dependent on the provided ground truth $F_0$.

These challenges have been addressed by methods based on self-supervised learning (SSL), which do not require labeled data.  Various SSL-based methods have been proposed~\cite{riou2023pesto,gfeller2020spice,choi2022nansy++}, among which one of the most notable is the methods that utilize pitch shift, exemplified by SPICE~\cite{gfeller2020spice}. These methods leverage the relative values of pitch shifts as the main objective, effectively bypassing the need for manually labeled data.
On the other hand, relative pitch difference is not the only clue for pitch prediction. Even when the target pitch labels are unavailable, clues for absolute pitch are also available by properly using DSP techniques. Leveraging these absolute pitch clues in the SSL-based pitch estimation schema is expected to further enhance the model performance.

Therefore, we propose SLASH, a novel pitch estimation model that integrates absolute pitch information into SSL-based model. SLASH incorporates absolute pitch in two ways leveraging DSP. The first approach involves using a DSP technique based on subharmonic summation (SHS)~\cite{ikemiya2016singing} to determine the prior distribution of the $F_0$. This distribution is used to regularize the estimation results, which encourage the model to incorporate the course $F_0$ structure into the pitch estimation.
The second approach optimizes the $F_0$ through gradient descent, using a loss between the target spectrogram and a spectrogram generated by a differentiable DSP (DDSP) module. Although this type of $F_0$ optimization is reported to be inefficient~\cite{torres2024unsupervised,engel2020ddsp}, the proposed method overcomes the problem by introducing a novel spectrogram generation method. Specifically, the method directly generates a spectrogram of periodic component from $F_0$, without generating a time-domain waveform. 
The direct generation of spectrogram from $F_0$ simplifies the backpropagation process, enabling the optimization of $F_0$ using gradient descent.
Furthermore, because the proposed method targets speech, it also aims to estimate the aperiodicity and voicing flags (V/UV) alongside the $F_0$. By passing the $F_0$, aperiodicity and spectral envelope to a DDSP synthesizer, the aperiodicity can also be optimized using gradient descent.

To demonstrate the effectiveness of the proposed method, experiments on pitch estimation for speech were conducted. Compared to DSP and SSL-based baselines, the proposed method showed superior results across all objective metrics, including $F_0$ RMSE and V/UV Error. Additionally, ablation studies confirmed that each component of the proposed method contributes to improved estimation accuracy.

\section{SLASH}
\subsection{Model overview}

\begin{figure*}[t]
    \centering
    \includegraphics[width=0.95\linewidth]{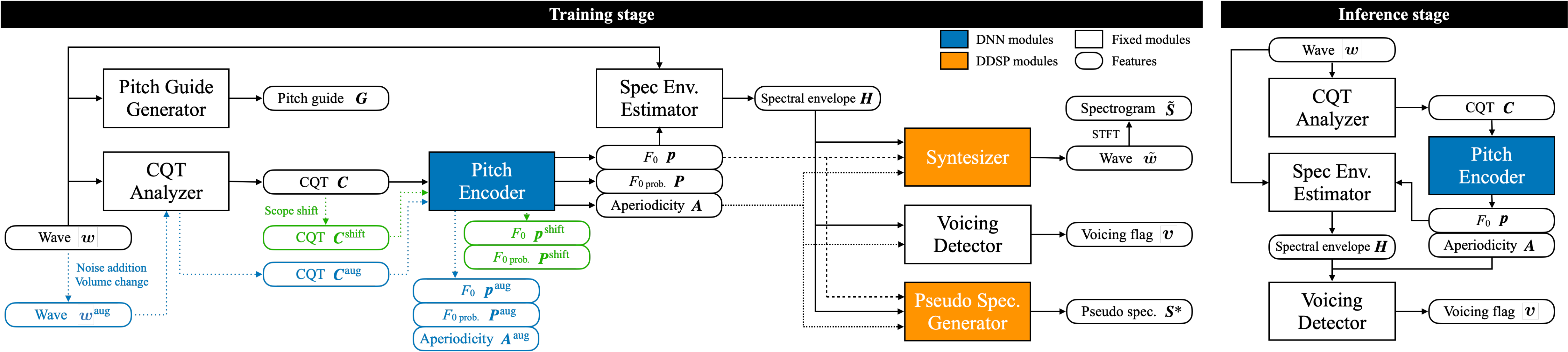}
    \vspace{-2mm}
    \caption{Model architecture of SLASH. The different types of arrows (dashed, dotted, and solid) are used solely for visibility purposes.}
    \label{fig:architecture}
    \vspace{-0mm}
\end{figure*}

Figure~\ref{fig:architecture} illustrates the architecture of SLASH.
As shown in the figure, only the Pitch Encoder is a DNN module, while all other modules are composed of DSP.
When viewed as a whole, SLASH functions as an analysis-synthesis system based on the source-filter model~\cite{kawahara2011technical, morise2016world, nakano2012spectral}. Specifically, the model decomposes the waveform $\bm{w}$ into $F_0$, spectral envelope, and aperiodicity\footnote{We call the group of $F_0$, spectral envelope, and aperiodicity as the \textit{vocoder features} for simplicity.} and then reconstructs the waveform $\tilde{\bm{w}}$ from them.
During training, spectral envelope is calculated by the Spec Env. Estimator, while $F_0$ and Aperiodicity are predicted by the Pitch Encoder. These features are then passed to the DDSP Synthesizer similar to~\cite{nercessian2022differentiable}, to reconstruct the waveform.
In addition, V/UV are estimated through the V/UV Detector. During inference, vocoder features and V/UV are estimated through these modules.
In the subsequent sections, we present a series of modules and methods aimed at enhancing prediction accuracy and stabilizing the training process.

\subsection{Pitch consistency loss}
Following the success in previous works~\cite{riou2023pesto, gfeller2020spice, choi2022nansy++}, we adopted the loss of pitch consistency. The key idea of this loss is to predict the pitch difference between the original and the pitch-shifted waveform to learn the relative pitch.
For training, Constant-Q Transform (CQT)~\cite{cqt} $\bm{C}\in \mathbb{R}^{T \times F_c}$ is first extracted from the input waveform $\bm{w}$, and then a pitch-shifted CQT $\bm{C}^{\text{shift}}$ is prepared by shifting the scope of $\bm{C}$. Here, $T$, and $F_c$ are the number of frames and the number of frequency bins in CQT, respectively. $\bm{C}$ and $\bm{C}^{\text{shift}}$ are fed into the Pitch Encoder to obtain two sequences of $F_0$ $\bm{p} \in \mathbb{R}^{T\times 1}$ and $\bm{p}^{\text{shift}}$, which are used to calculate the pitch consistency loss defined as follows:
\begin{align}
L_{\text{cons}} = \frac{1}{T} \sum_{t=1}^{T} h\left( \left| \log_2 \bm{p}_t- \log_2 \bm{p}^{\text{shift}}_t+ \frac{d}{12} \right| \right),
\end{align}
where $ h(\cdot) $ denotes the Huber norm~\cite{Huber1964}.
$d$ is the frequency axis shift in semitones, converted to octaves by dividing by 12.
The architecture of the Pitch Encoder is based on the Pitch encoder of~\cite{choi2022nansy++}, with slight modifications in the output. Specifically, unlike the original module that predicts the amplitudes of periodic and aperiodic components, the Pitch Encoder in SLASH outputs the probability matrix of $F_0$ $\bm{P}\in \mathbb{R}^{T \times F}$ and an $b$-dimensional Band Aperiodicity (BAP) $\bm{B} \in \mathbb{R}^{T \times b}$. Here, $F$ denotes the number of frequency bins. $\bm{p}$ is calculated by taking the weighted average of $\bm{P}$. 
Aperiodicity $\bm{A}\in \mathbb{R}^{T \times K}$ is then calculated by linearly interpolating $\bm{B}$ on the logarithmic amplitude, where $K$ represents the number of frequency bins.
For $\bm{C}^{\text{shift}}$, $\bm{P}^{\text{shift}}$ is predicted accordingly.

\subsection{Pitch guide}
\begin{figure}[t]
    \centering
    \includegraphics[width=1.0\linewidth]{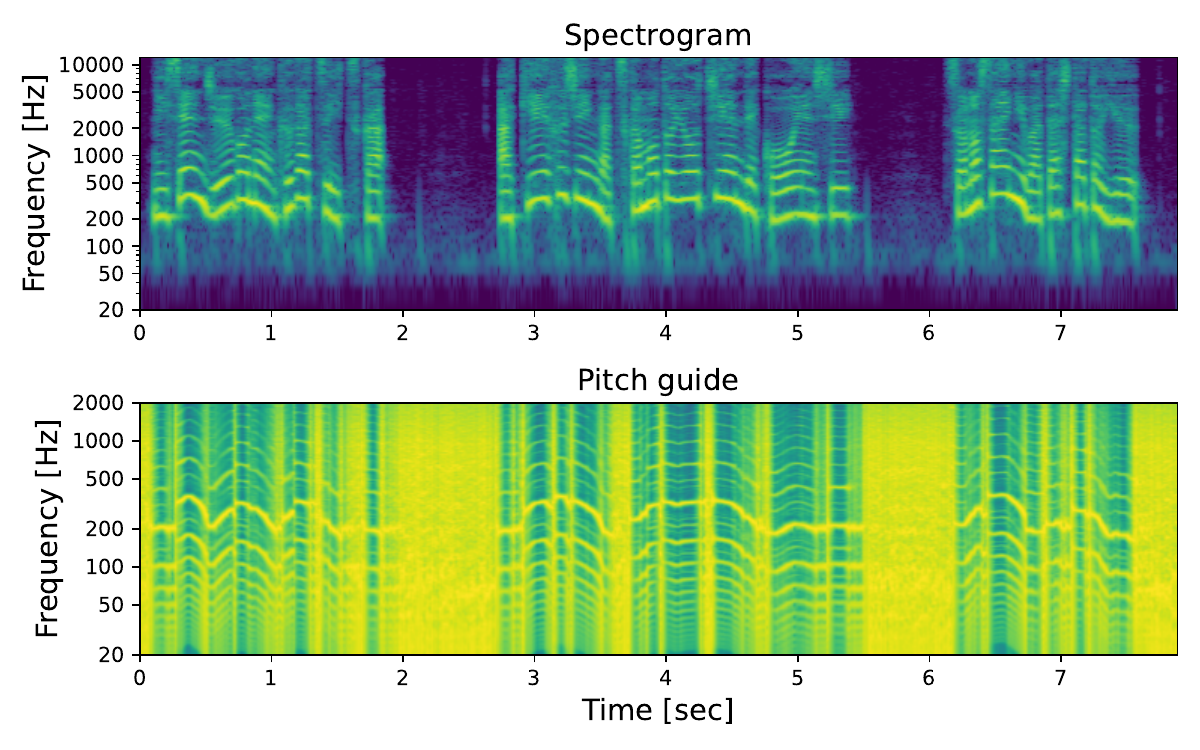}
    \vspace{-6mm}
    \caption{An example of pitch guide $\bm{G}$ with log spectrogram.}
    \label{fig:pitch_guide}
    \vspace{-0mm}
\end{figure}

Although the pitch consistency loss is effective to learn the relative pitch, it is difficult to predict the absolute pitch with this loss alone. To address this issue, we introduce a pitch guide $\bm{G} \in \mathbb{R}^{T \times F}$, which is a frame-level prior distribution of $F_0$ calculated by the DSP-based Pitch Guide Generator.  
$\bm{G}$ is obtained through the following steps. 1) The fine structure spectrum in logarithmic domain $\psi(\bm{S})\in \mathbb{R}^{T\times K}$ is calculated as:
\begin{equation}
    \psi(\bm{S}) = \log(\bm{S}) - W(\log(\bm{S})), \label{eq:psi}
\end{equation}
where $\bm{S} \in \mathbb{R}^{T \times K}$ denotes the input amplitude spectrogram, and $W(\cdot)$ represents the calculation of the spectral envelope by lag-window method~\cite{tohkura1978spectral}. 2) SHS~\cite{ikemiya2016singing} is applied to $\exp(\psi(\bm{S}))$, to obtain $\bm{G}' \in \mathbb{R}^{T \times F}$. Note that the frequency scale is changed from $K$ to $F$ in this process. 3) $\bm{G}'$ is normalized so that the maximum of every time frame equals 1, resulting in $\bm{G}$.
Figure~\ref{fig:pitch_guide} shows an example of pitch guide $\bm{G}$ with the corresponding spectrogram. As shown in the figure, the pitch guide well captures the harmonic structure of speech in voiced region, and is expected to serve as a clue for determining the absolute pitch. 
To encourage the Pitch Encoder to use $\bm{G}$ as a reference of absolute pitch, the pitch guide loss is defined as follows:
\begin{align}
L_{\text{g}} = \frac{1}{T} \sum_{t=1}^{T} \max\left(1 - \sum_{f} \bm{P}_{t,f} \cdot \bm{G}_{t,f} - m, 0\right),
\label{eq:L_guide}
\end{align}
where $ \bm{P}_{t,f} $ represents $\bm{P}$ at time frame $t$ and frequency $f$, and $ m $ is the hinge parameter to relax the constraint.
Pitch guide loss is also applied to $\bm{P}^{\text{shift}}$. In this case, the pitch guide is shifted according to the amount of pitch shift $\Delta f$:
\begin{align}
L_{\text{g-shift}} = \frac{1}{T} \sum_{t=1}^{T} \max\left( 1 - \sum_{f} \bm{P}^{\text{shift}}_{t,f} \cdot \bm{G}_{t,f-\Delta f} - m, 0 \right). \notag
\end{align}

\vspace{-2mm}
\subsection{$F_0$ optimization with gradient descent}
Another approach to learn the absolute pitch is to synthesize the target waveform from vocoder features through DDSP or DNNs and update the input $F_0$ through gradient descent~\cite{andrychowicz2016learning}. However, the gradient descent of $F_0$ from waveform is reported to be ineffective due to the limited perception of gradient orientation relative to $F_0$, and the presence of numerous local minima~\cite{torres2024unsupervised,engel2020ddsp}. To overcome this issue, we propose a novel spectrogram generation method, which directly generate the spectrogram of periodic component from $F_0$, bypassing the waveform generation. This approach is expected to stabilize $F_0$ back propagation by avoiding complex waveform generation processes.
Particularly, the $F_0$ gradient descent is performed through DDSP-based Pseudo Spec. Generator, which receives predicted vocoder features as input. The gradient descent is performed by the following steps.

\noindent\textbf{Excitation spectrogram generation:} The Pseudo Spec. Generator generates a pseudo spectrogram of the periodic excitation signal $\bm{E}^{*}_{\text{p}} \in \mathbb{R}^{T \times K}$ from $\bm{p}$. First, the phase of the signal at time $t$ is calculated using the $F_0$ $\bm{p}_t$, sampling rate $f_s$, and frequency index $k \in \{1, 2, \ldots, K\}$, and $K$:
\begin{align}
     \Phi_{t,k}= \frac{f_s}{2\bm{p}_tK}k. \notag
\end{align}
Then, a triangle wave oscillating $\bm{X}$ between $-1$ and $1$ is generated based on $\Phi_{t,k}$:
\begin{align}
    \bm{X}_{t,k} = 
    \begin{cases} 
    -1, & \text{if } \Phi_{t,k} < 0.5 \\
    4 \left| \Phi_{t,k} - \left\lfloor \Phi_{t,k} \right\rfloor - 0.5 \right| - 1, & \text{otherwise}, \notag
    \end{cases}
\end{align}
where $\lfloor \cdot \rfloor$ denotes the floor function. The pseudo periodic excitation spectrogram $\bm{E}^*_{\text{p}}$ is then obtained as:
\begin{align}
     \bm{E}^{*}_{\text{p}} = \max(\bm{X}, \varepsilon)^2 + \left|\bm{Z} \cdot \varepsilon\right|,
\label{eq:E_p}
\end{align}
where $\varepsilon$ denotes a small magnitude, and $\bm{Z}\in \mathbb{R}^{T \times K}$ denotes a matrix with each element being independently drawn from standard normal distribution.

\noindent\textbf{Loss calculation:}
After $\bm{E}^*_{\text{p}}$ is obtained, the entire pseudo spectrogram $\bm{S}^*\in \mathbb{R}^{T \times K}$ is calculated as follows:
\begin{align}
    \bm{S}^{*} = (\bm{E}^*_{\text{p}} \odot \bm{H} \odot (1 - \bm{A})) + \left( \mathcal{F}(\bm{e}_\text{ap}) \odot \bm{H} \odot \bm{A} \right), \label{eq:s_star}
\end{align}
where $\odot$ denotes the Hadamard product, $\bm{H}\in \mathbb{R}^{T \times K}$ denotes the spectral envelope, $\bm{e}_\text{ap}$ represents the aperiodic excitation signal in time-domain, and $\mathcal{F}$ represents the Fourier transform.  The term $\mathcal{F}(\bm{e}_\text{ap}) \odot \bm{H} \odot \bm{A}$ represents the spectrogram of aperiodic component. It is obtained by applying an STFT to the aperiodic component of the Synthesizer output $\tilde{\bm{w}}$
\footnote{\label{note:synth} In practice, the periodic and aperiodic component of $\tilde{\bm{w}}$ is generated from $\bm{e}_{\text{p}}$ and ${\bm{e}_\text{ap}}$ in \textbf{time}-domain using minimum-phase response. The frequency representation is used in eq. (\ref{eq:s_star}) and (\ref{eq:s_tilde}) for brevity.}.

\begin{figure}[t]
    \centering
    \includegraphics[width=1.0\linewidth]{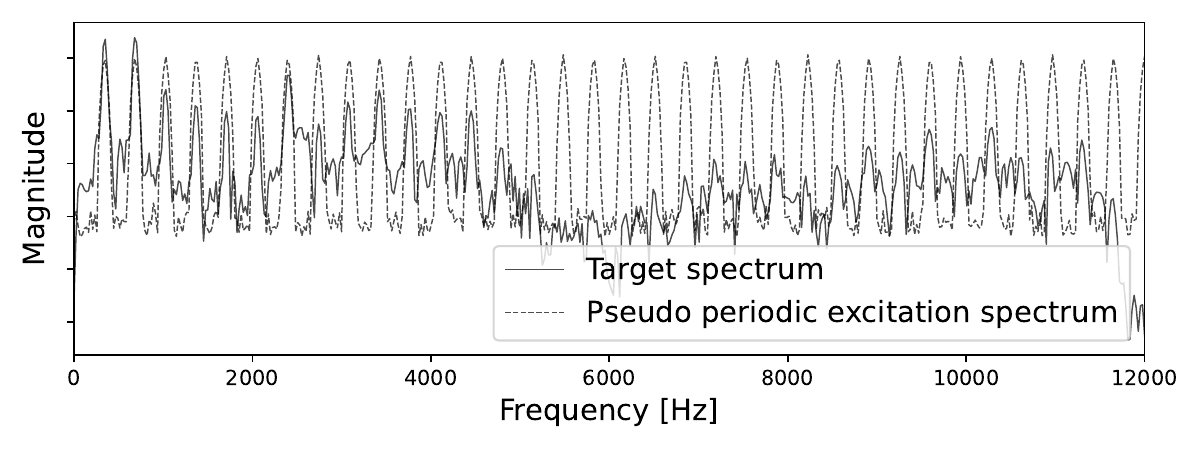}
    \vspace{-6mm}
    \caption{An example of the target spectrum $\bm{S}_t$ and the pseudo periodic excitation spectrum $\bm{E}^*_t$ generated from optimized $\bm{p}_t$.}
    \label{fig:Pseudo_mag}
    \vspace{-0mm}
\end{figure}

Finally, the loss is defined as an L1 norm of the predicted and the target log fine structure spectrum:
\begin{equation}
    L_{\text{pseudo}} = \left\lVert \psi(\bm{S}^*) - \psi(\bm{S}) \right\rVert_1 \odot (\bm{v} \times \bm{1}_K), \label{eq:L_pseudo}
\end{equation}
and $\bm{v} \in \mathbb{R}^{T\times1}$ is a mask of V/UV, which will be introduced in eq. (\ref{eq:vuv}).
This loss aims to align the peak positions of the harmonics in the excitation signal spectrum $\bm{E}^*_t$ with those in the target spectrum $\bm{S}_t$. Figure~\ref{fig:Pseudo_mag} shows the case where $\bm{p}_t$ is perfectly optimized. In this case, we can see that the peak positions of $\bm{E}^*_t$ and $\bm{S}_t$ are perfectly aligned.
Note that the computational graphs of features other than $\bm{p}$ were detached to focus on the $F_0$ prediction.

\subsection{Aperiodicity prediction}
Since SLASH focuses on speech signals, aperiodicity $\bm{A}$ and V/UV $\bm{v}$ are predicted along with the $F_0$ $\bm{p}$. $\bm{A}$ is optimized through the gradient descent using the loss between the target spectrogram $\bm{S}$ and another spectrogram $\tilde{\bm{S}}\in \mathbb{R}^{T \times F}$, which is obtained from $\tilde{\bm{w}}$ through STFT.
In the frequency domain, \(\tilde{\bm{S}}\) can be expressed by the following equation\footref{note:synth}:
\begin{equation}
\tilde{\bm{S}} = (\mathcal{F}(\bm{e}_\text{p}) \odot \bm{H} \odot (1 - \bm{A})) + (\mathcal{F}(\bm{e}_\text{ap}) \odot \bm{H} \odot \bm{A}), \label{eq:s_tilde}
\end{equation}
where $\bm{e}_\text{p}$ represents a time-domain periodic excitation signal, created by combining multiple sine waves based on $\bm{p}_t$ and its harmonics~\cite{nercessian2022differentiable}.
Instead of the simple L1 norm, a generalized energy distance (GED)~\cite{gritsenko2020spectral} is used to stabilize optimization:

\begin{equation}
L_{\text{recon}} = \left\lVert \psi(\tilde{\bm{S}}^1) - \psi(\bm{S}) \right\rVert_1 - \alpha \left\lVert \psi(\tilde{\bm{S}}^1) - \psi(\tilde{\bm{S}}^2) \right\rVert_1 ,
\label{eq:L_recon}
\end{equation}
where $\tilde{\bm{S}}^1$ and $\tilde{\bm{S}}^2$ denote two different generated spectrograms, and $\alpha$ is a weight parameter that controls the influence of the repulsive term in the GED. Similar to $L_{\text{pseudo}}$, the loss is calculated by focusing on the fine structure spectrum. 

The Voicing Detector calculates the V/UV $\bm{v}$ as:
\begin{equation}
    \bm{v}' = \frac{M_{\text{p}}}{M_{\text{p}} + M_{\text{ap}}}, 
    \bm{v} = 
\begin{cases} 
1, & \text{if } \bm{v}' \geq \theta \\
0, & \text{otherwise}.
\end{cases}
\label{eq:vuv}
\end{equation}
Here, $M_{\text{p}}$ and $M_{\text{ap}}$ are the magnitude of periodic and aperiodic components, which are calculated as the weighted sum of spectral envelope $\bm{H}$ by $\bm{A}$. $\theta$ is a threshold parameter.

\subsection{Noise robust training \label{sec:noise}}
To enhance the robustness to noisy inputs, we prepare augmented waveforms by randomly adding noise and changing the volume for input waveforms~\cite{riou2023pesto}. $\bm{C}^{\text{aug}}$ is calculated through the CQT Analyzer, and $\bm{p}^{\text{aug}}$, $\bm{P}^{\text{aug}}$, and $\bm{A}^{\text{aug}}$ are obtained through the Pitch Encoder. Three losses are defined for these features. The first loss $L_{\text{aug}}$ is similar to $L_{\text{cons}}$, which is defined as the Huber norm between $\bm{p}$ and $\bm{p}^{\text{aug}}$. The second loss $L_{\text{g-aug}}$ is almost the same as $L_{\text{g}}$, except that $\bm{P}$ is substituted with $\bm{P}^{\text{aug}}$. These losses are expected to enhance the robustness of pitch estimation against noisy inputs. The last loss $L_{\text{ap}}$ is introduced to enhance the noise robustness of aperiodicity prediction:
\begin{equation}
L_{\text{ap}} = \left\lVert \log(\bm{A}^{\text{aug}}) - \log(\bm{A}) \right\rVert_1.\nonumber
\label{L_ap}
\end{equation}

\section{Experiments}

\begin{table*}[t]
\caption{Evaluation results of SLASH compared to digital signal processing, self-supervised learning, and supervised learning baselines. \textbf{Bold} font denotes the best score among models except supervised learning model, which is \textcolor{gray}{grayed out}.}
\label{tbl:eval}
\begin{center}
\vspace{-12pt}
\scalebox{0.99}{
    \begin{tabular}{l|c|c|c|c|c}\hline
    \multirow{2}{*}{Model} & \multirow{2}{*}{Trained on} & \multicolumn{3}{c|}{Pitch} & \multicolumn{1}{c}{V/UV} \\\cline{3-6}
                           &             & {RPA (RCA) 50c ↑}      & {RPA (RCA) 100c ↑}     & {log-$F_0$ RMSE ↓}& {V/UV ER ↓}     \\\hline
    {DIO}                  & -           & 0.943 (0.943)          & 0.976 (0.976)          & 0.030          & 0.052          \\
    {Harvest}              & -           & 0.945 (0.947)          & 0.972 (0.974)          & 0.045          & 0.083          \\\hline
    \multirow{2}{*}{SLASH} & LibriTTS-R  & \textbf{0.969 (0.969)} & \textbf{0.990 (0.990)} & 0.018          & \textbf{0.033} \\
                           & MIR-1K      & 0.967 (0.967)          & \textbf{0.990 (0.990)} & \textbf{0.017} & \textbf{0.033} \\\hline
    {PESTO}                & MIR-1K      & 0.962 (0.966)          & 0.982 (0.986)          & 0.057          & 0.098          \\
    \textcolor{gray}{CREPE full} & \textcolor{gray}{many (supervised)} & \textcolor{gray}{0.978 (0.978)} & \textcolor{gray}{0.989 (0.990)} & \textcolor{gray}{0.038} & \textcolor{gray}{0.101} \\\hline
    \end{tabular}
}
\vspace{-8pt}
\end{center}
\end{table*}

\subsection{Experimental setup}
\noindent\textbf{Datasets:}
In our experiments, we used two datasets.
The first one is \textbf{LibriTTS-R}~\cite{koizumi2023libritts}, a high-quality multi-speaker English corpus of 585 hours of read English speech designed for text-to-speech use.
The second one is \textbf{MIR-1K}~\cite{hsu2009improvement}, consisting of 1,000 song clips extracted from 110 karaoke songs. It contains both vocal and instrumental tracks and provides pitch and voicing flag annotations for the vocals.
For the evaluation of all systems, we used 250 randomly selected phrases from the MIR-1K dataset.
During the experiments, all data were resampled to a sampling rate of 24 kHz.
 
\noindent\textbf{Model details:}
The CQT was configured with a frame shift of 5 ms, a minimum frequency $f_{\text{min}}$ of 32.70 Hz, and a total of 205 bins, with 24 bins per octave. To accommodate rapid pitch changes, the filter scale was set to 0.5. The pitch encoder processes the central 176 bins. The shift range for $\bm{C}^{\text{shift}}$ is set to $\pm14$ bins. $\bm{C}^{\text{aug}}$ was computed by applying random noise (maximum SNR of -6 dB) and volume variations ($\pm6$ dB) to the input waveform.
The pitch encoder is designed to output a 1024-dimensional $F_0$ probability distribution (i.e. $F=1024$) mapped onto a logarithmic frequency scale ranging from 20 Hz to 2 kHz, along with an 8-dimensional BAP.
The parameter $m$ in eq. (\ref{eq:L_guide}) was set to 0.5, $\alpha$ in eq. (\ref{eq:L_recon}) was set to 0.1, $\theta$ in eq. (\ref{eq:vuv}) was set to 0.5, and $\varepsilon$ in eq. (\ref{eq:E_p}) was set to 0.001.
The weights for the core losses were set as follows: $L_{\text{cons}}$ and $L_{\text{pseudo}}$ were set to 10.0, $L_{\text{recon}}$ was set to 5.0. Additionally, $L_{\text{g}}$, $L_{\text{g-shift}}$, $L_{\text{aug}}$, $L_{\text{g-aug}}$, and $L_{\text{ap}}$ were set to 1.0.
During the training, we utilized a dynamic batching~\cite{hayashi2020espnet} with an average batch size of 17 samples to create mini-batches. The model was trained for 100,000 steps using the AdamW optimizer~\cite{los2019adamw} with a learning rate of 0.0002.

\noindent\textbf{Comparison models:}
We prepared two SLASH models trained on LibriTTS-R and MIR-1K. The latter was used to better align with the baseline model conditions.
A total of 700 phrases were used for training and 50 phrases for validation.
As the comparison models, we used methods based on DSP, SSL, and supervised learning.
DIO~\cite{morise2009fast} and Harvest~\cite{morise2017harvest} are DSP-based $F_0$ estimation methods implemented in the WORLD~\cite{morise2016world} speech analysis-synthesis system.
For sections where the estimated $F_0$ by DIO and Harvest is 0 Hz, linear interpolation of $F_0$ was performed on a logarithmic scale.
PESTO is an SSL-based baseline model. We used the official pre-trained model of PESTO\footnote{https://github.com/SonyCSLParis/pesto}, which is trained on the MIR-1K dataset.
CREPE~\cite{kim2018crepe} is a model using supervised learning. We used the official pre-trained model with Viterbi smoothing\footnote{https://github.com/marl/crepe}, which is trained on various datasets\cite{hsu2009improvement, duan2010multiple, bittner2014medleydb, mauch2014pyin, salamon2017analysis, engel2017neural}, including MIR-1K.
Note that CREPE uses some ground-truth pitch labels during training, while other models do not. For the V/UV estimation of PESTO and CREPE, we used confidence values.
Frames with a confidence level of 0.5 or higher were classified as voiced.

\noindent\textbf{Evaluation metrics:}
We predicted pitch and V/UV for the test set of MIR-1K, and evaluated the results with several metrics.
Raw pitch accuracy (RPA) measures the proportion of voiced frames where the estimated pitch is within 50 cents of the true pitch. We additionally consider cases within 100 cents~\cite{salamon2014melody}.
Raw chroma accuracy (RCA) accounts for the proportion of RPA that allows for octave errors~\cite{salamon2014melody}.
The log-$F_0$ RMSE represents the root mean square error between the true and estimated $F_0$ in logarithmic domain, calculated over voiced frames.
Lastly, the voiced/unvoiced error rate (V/UV ER) indicates the error rate in V/UV classification.

\subsection{Results}

\subsubsection{Clean signals}
Table~\ref{tbl:eval} summarizes the evaluation results on MIR-1K. 
Among the DSP and SSL-based methods, which do not use the target labels during training, SLASH achieved the best scores in all metrics. It also showed competitive performance against CREPE, which uses the ground-truth pitch labels in MIR-1K.
These results confirm the high pitch and V/UV estimation performance of SLASH, which takes the advantage of both SSL and DSP. 
In terms of the metrics, SLASH particularly performed well for the log-$F_0$ RMSE compared to the RPA (RCA). This result suggests that the prediction of SLASH rarely deviates significantly from the true value even when it does not fall within the accuracy range of RPA.
In addition, SLASH outperforms the baseline SSL model PESTO in all metrics. This suggests that introducing loss functions for absolute pitch is effective to enhance the pitch prediction accuracy.
In terms of the training dataset, SLASH trained on LibriTTS-R demonstrated performance nearly equivalent to that trained on MIR-1K, although the former is an out-of-domain dataset. This indicates that training SLASH on a large amount of dataset like LibriTTS-R leads to a good generalization performance to out-of-domain dataset such as singing voice.

\subsubsection{Robustness to noise}
\begin{figure}[tb]
    \centering
    \includegraphics[width=1.0\linewidth]{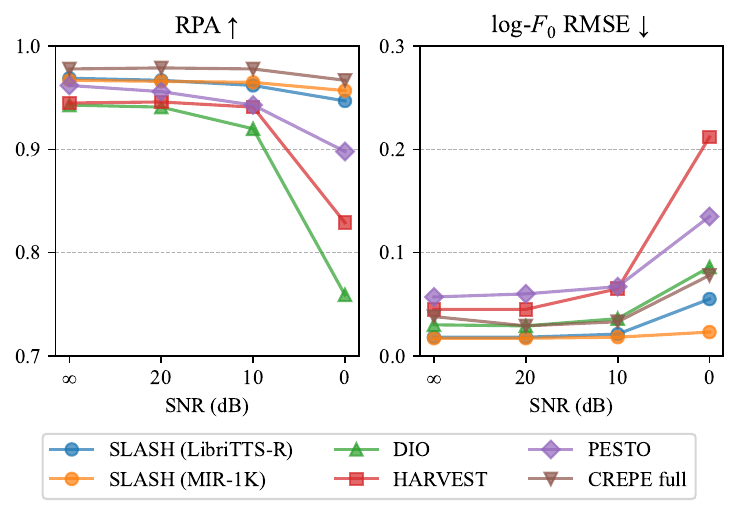}
    \vspace{-6mm}
    \caption{RPA and log-$F_0$ RMSE in different SNR conditions.}
    \label{fig:noise}
    \vspace{-0mm}
\end{figure}

Figure~\ref{fig:noise} illustrates the RPA and log-$F_0$ RMSE under conditions with added white noise. The results demonstrate that SLASH exhibits high robustness even in noisy environments. Notably, at an SNR of 0 dB, SLASH experienced minimal deterioration, while other systems showed significant degradation in log-$F_0$ RMSE. This is likely due to the waveform augmentation and related loss functions as described in Sec.\ref{sec:noise}.

\subsubsection{Ablation study}

Table~\ref{tbl:ablations} shows the results of the ablation study.
Without $L_{\text{g}}$ and $L_{\text{pseudo}}$ (i.e., without absolute pitch losses), the scores significantly worsened.
This is likely because an absolute reference for pitch is not adequately provided and the model was unable to predict the absolute pitch.
Without $L_{\text{pseudo}}$, i.e., when $L_g$ is added, RPA and log-$F_0$ RMSE significantly improve, but not to the same extent as SLASH. These results suggest that while $L_{\text{g}}$ helps capture overall pitch trends, $L_{\text{pseudo}}$ also plays an essential role in enhancing the pitch prediction accuracy.
Finally, w/o GED, $L_{\text{ap}}$ represents the case where the aperiodicity is optimized by a simple L1 norm in eq. (\ref{eq:L_recon}). In this case, the V/UV ER worsened, which confirms the necessity of GED and aperiodicity regularization. In addition, other pitch objectives are also worsened. This is likely due to the use of unreliable V/UV in the calculation of $L_{\text{pseudo}}$ (as shown in eq. (\ref{eq:L_pseudo})).

\begin{table}[t]
\caption{
    Evaluation results of the ablation study.
}
\label{tbl:ablations}
\begin{center}
\vspace{-12pt}
\scalebox{0.99}{
    \begin{tabular}{l|c|c|c}\hline
    \multirow{2}{*}{Model} & \multicolumn{2}{c|}{Pitch} & \multicolumn{1}{c}{V/UV} \\\cline{2-4}
                                                & {RPA ↑}   & {log-$F_0$ RMSE ↓} & {V/UV ER ↓} \\\hline
    SLASH                                       & 0.969     & 0.018              & 0.033      \\\hline
    w/o $L_{\text{g}}$, $L_{\text{pseudo}}$     & 0.000     & 0.768              & 0.703      \\
    w/o $L_{\text{pseudo}}$                     & 0.934     & 0.031              & 0.037      \\
    w/o GED, $L_{\text{ap}}$                    & 0.913     & 0.187              & 0.243      \\\hline
    \end{tabular}
}
\end{center}
\end{table}

\vspace{-5pt}
\section{Conclusion}

We proposed SLASH, an SSL-based pitch estimation method with DSP-derived absolute pitch information. By incorporating the absolute pitch into the model, SLASH enhanced the pitch prediction accuracy of conventional SSL-based methods, which depend on relative pitch objectives. Future work includes expanding SLASH to real-time pitch estimation.

\bibliographystyle{IEEEtran}
\bibliography{mybib}

\end{document}